\documentclass[fleqn,10pt]{wlscirep}
\title{Voltage assisted asymmetric nanoscale wear on ultra-smooth diamond like carbon thin films at high sliding speeds}

\author[1*]{Sukumar Rajauria}
\author[1]{Erhard Schreck}
\author[1]{Bruno Marchon}
\affil[1]{HGST, a Western Digital Company, Recording Sub System Staging and Research, San Jose, CA 95135 USA.}

\affil[*]{corresponding.sukumar.rajauria@hgst.com}

\begin{abstract}
\textbf{The understanding of tribo- and electro-chemical phenomenons on the molecular level at a sliding interface is a field of growing interest. Fundamental chemical and physical insights of sliding surfaces are crucial for understanding wear at an interface, particularly for nano or micro scale devices operating at high sliding speeds. A complete investigation of the electrochemical effects on high sliding speed interfaces requires a precise monitoring of both the associated wear and surface chemical reactions at the interface. 
Here, we demonstrate that head-disk interface inside a commercial magnetic storage hard disk drive provides a unique system for such studies. The results obtained shows that the voltage assisted electrochemical wear lead to asymmetric wear on either side of sliding interface.} 

\end{abstract}
\begin{document}

\flushbottom
\maketitle
\thispagestyle{empty}

\section*{Introduction}

Wear is broadly classified in two categories: physical wear and progressive wear. Physical wear is further classified as adhesive, fatigue and abrasive wear which all lead to the formation or transfer of material across the sliding interface \cite{Dowson,Wimmer03,Mosey05Science,Bhushan94Nature,Gosvami15Science,SukumarAPL15}. At macro or micro scale it follows Archard's law describing fracture and plastic deformation with wear volume being proportional to both the applied load and sliding distance \cite{ArchardJAP53,JiaWear97,ChungTL03}. Such phenomena lead to catastrophic wear and are rare in nanoscale devices where the progressive chemically assisted  wear is likely to be dominant. Recently observed nanoscale wear showed that stress-assisted chemical reactions occur through an atom-by-atom process \cite{GneccoPRL02,GotsmannPRL08,BhaskaranNatureNano10,JacobsNatureNano13, HanggiRMP90}. In chemical assisted wear, environmental species like oxygen and humidity play a critical role. For instance, oxygen can readily
chemisorb on carbon surfaces, leading to surface oxides such as carbonyl groups, thereby depleting the carbon surface during thermal desorption \cite{Robertson02,MarchonCarbon88,Marchon90IEEE,StromASME91,DaiIEEE03,KonicekPRB08,Marino11Lagmuir}. In addition to environmental species, the chemical potential across the sliding interface is expected to affect the surface chemical reaction rates. On the macroscale level many studies were conducted to understand the impact of chemical potential on surface oxidation \cite{KinoshitaCarbon73,GallagheraPCCS09,ZilibottiPRB09}, but at the nanoscale a quantitative understanding of its mechanism and impact on wear appears to be little understood.

A complete investigation of the electrochemical effects on high sliding speed interfaces requires a precise monitoring of both the associated wear and surface chemical reactions at the interface. The head-disk interface inside a commercial magnetic storage hard disk drive provides a unique system for such studies. In a hard disk drive, the head has an embedded micro-scale heater which produces through thermal expansion a well-defined mechanical protrusion. This allows to adjust the head-to-disk interference level within sub-nanometer precision over a contact area of several square micrometers. The relative sliding speed between head and disk ranges from 10-40 $m/s$. It is worth noticing here that while the vertical spacing is of the same order of various AFM based studies, the sliding speed is nearly six orders of magnitude higher, thus allowing a unique set-up for a systematic study of nanoscale wear at high sliding speeds. Interfering surfaces of the head and the disk are coated with an amorphous diamond like carbon which has exceptional mechanical properties like low friction and wear rate \cite{Mate87PRL,Robertson01TFS,Ferrari04Surface}. 

In this letter, we report the precise monitoring of both the carbon overcoat wear and the corresponding interfacial current on the nanoscale high sliding speed interface. Carbon overcoat wear is monitored and calibrated using the embedded micro-heater power. The interface current between the head and the disk monitor the rate of electrochemical oxidation of the carbon overcoat. We show that the interface current decay sharply with time indicating the chemical passivation of the surface carbon dangling bonds that are created while sliding. This unique approach provides an in-depth understanding of the electrochemically assisted wear at high sliding speeds which until now has never been applied to such devices.

\section*{Results}

A typical head-disk interface setup features the head flying on top of the disk similar to the one studied in Ref \cite{SuhTL06,SukumarAPL15}. The disk is fabricated by depositing a magnetic multilayer film structure onto a glass substrate, then coated with 3 $nm$ amorphous nitrogenated carbon (protective overcoat layer), and finally covered with a molecular layer of perfluoropolyether polymer lubricant ($\sim$ 1 $nm$ thick). The electrically conductive ceramic substrate head is also coated with 1.4 $nm$ of diamond like carbon on top of a 0.3 $nm$ Silicon based adhesive layer, making the head and the disk interface a carbon-lubricant-carbon sliding interface. The disk facing head surface is carefully shaped through etching such that while flying on top of the disk an airbearing lift force is generated that keeps it afloat over the disk in the nanometer range \cite{cml,ZhengTL10}. The linear sliding speed in the described experiments is set to 10 $m/s$. The initial clearance (physical gap) between the head and the disk is typically 10 $nm$. Clearance is controlled precisely using the embedded micro-heater in the head \cite{ZengIEEE11,CanchiAT12}. We estimate the wear depth by continuous monitoring of the micro-heater power and later verified the wear using AFM, scanning electron microscope (SEM) and Auger electron microscopy (Auger) \cite{ChenTL14}. Contact between head and disk is detected using an acoustic emission (AE) sensor \cite{BhushanIEEE03}.

%******************************************
\begin{figure*}[htbp]
\begin{center}
\includegraphics[width=7 in]{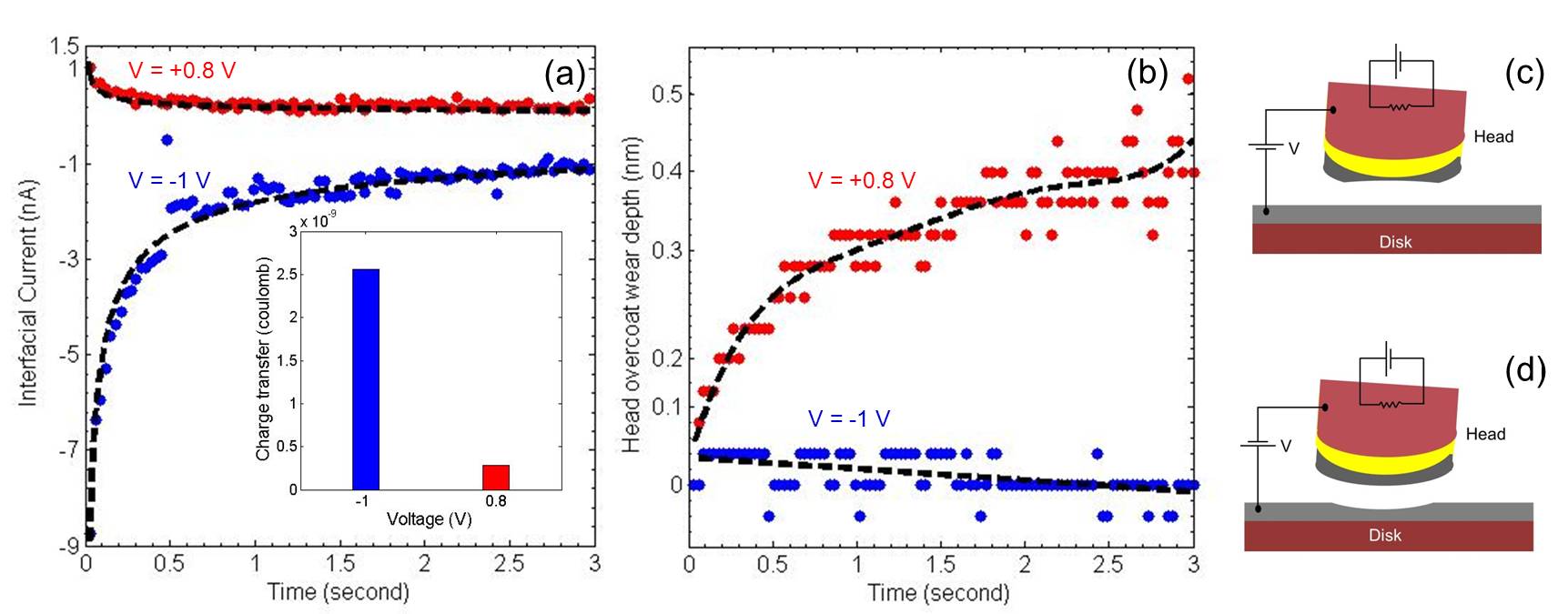}
\end{center}
\caption {\textbf{Simultaneous measurement of interfacial current and overcoat wear:} (a) Blue and Red dots represent the rapid decay of the interfacial current between the head and the disk with -1 $V$ and +0.8 $V$ applied on head. Dashed line is a fit to Eq. 1. Inset shows the associated charge transfer for the two polarities. (b) Blue and Red dots show the simultaneous in-situ measurement of overcoat wear depth as a function of time. (c), (d) Shows the schematic cartoon of the wear on head and disk overcoat for positive and negative voltage on the head. Clearance between head and disk is controlled precisely using the embedded micro-heater in the head. }
    \label{fig:1}
\end{figure*}
%%******************************************************************************************************

While in operation, a voltage is applied directly to the head overcoat while the disk was electrically grounded. Voltage difference across the interface decrease the initial flying clearance, which is symmetric to 0 $V.$ Variation in initial flying clearance across various applied voltages here is estimated to be around 3$\%$ of the initial flying clearance (see supplementary Figure S1) and part-to-part variation in initial flying clearance is estimated to be around 5$\%$. Interfacial current was measured across the interface. The load is set using the  micro-heater design. Air bearing simulations estimate the normal load to increase by 0.25 $mN/mW$ of excess heater power (Air bearing simulation using HGST internal code). Uncertainty due to variation in initial flying clearance is estimated to change the contact force by 10$\%$. Excess heater power is defined as excess power applied to the micro-heater after head-to-disk contact was detected. Figure 1a shows the interfacial current at two applied voltages of +0.8 $V$ and -1 $V$ on two heads under a normal load condition of 2.5 $mN$. For both polarities, the interfacial current at head disk interface decay with time. For further insight, an in-situ monitoring of the wear is desired to gain better insight into the impact of electrochemical oxidation during sliding. The head disk interface has a unique feature with the micro-heater power calibrated precisely to measure the head overcoat wear depth in a continuous manner during the experiment (see Supplementary Information for more detail) \cite{SukumarAPL15}. Figure 1b shows the wear depth profile of the head carbon overcoat for the respective voltage condition as a function of time. For positive voltage, the 0.3 nm thin carbon overcoat wears out within 3 second of intimate head disk contact (see red dots in Figure 1b).

%%******************************************************************************************************
\begin{figure}[htbp]
\begin{center}
\includegraphics[width=3.6 in]{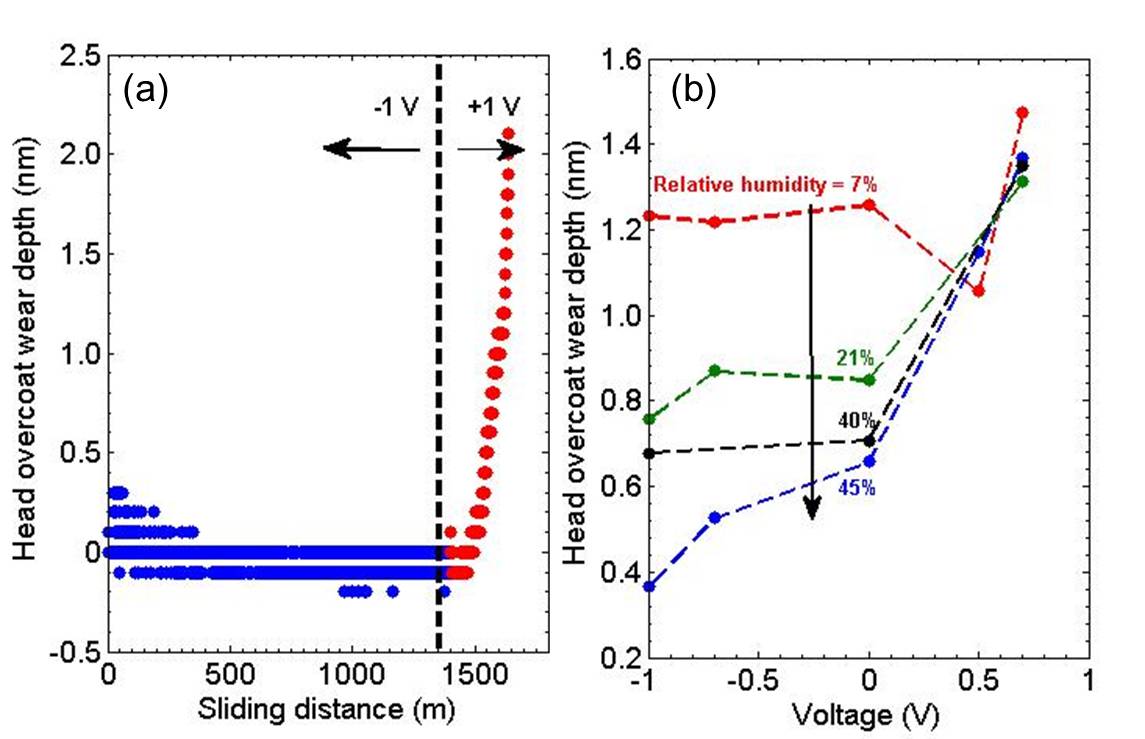}
\end{center}
\caption {\textbf{Head overcoat wear depth:} (a) Shows the wear depth profile as a function of sliding distance under an applied bias of -1 V and +1 V on the head overcoat. (b) Shows the wear depth profile of the head overcoat wear as a function of interfacial voltage on the head under different environment conditions.}
    \label{fig:2}
\end{figure}
%******************************************************************************************************

We associate the rapid decay in current to the electrochemical oxidation of the carbon overcoat, with voltage polarity governing whether the head or disk is undergoing oxidation. It is worth mentioning that similar rapid decay in current has previously been observed in macroscopic systems such as the carbon based fuel cell \cite{KinoshitaCarbon73,GallagheraPCCS09}. In an acidic environment, the current across the fuel cell decays rapidly with time across all potentiostatics. The electrochemical oxidation of carbon in fuel cell is written as: ${C + 2H_{2}O \to CO_{2} + 4H^{+} + 4e^{-}}$. Empirically, the measured oxidative electrochemical current is adequately represented by a simple power law expression:
\begin{equation}
i=kt^{-n} + i_{o} 
\label{eq:1}
\end{equation}
where $t$ is the time, $i$ is the specific current, $k$ is the rate parameter which is a function of both the temperature and potential, $n$ is the time decay exponent, and $i_{o} $ is the ohmic current. This decay in current with time is mainly attributed to two competing parallel reactions involving one to passivate the surface and another to oxidize the carbon producing carbon dioxide CO$_{2}$.  Eq. (1) fits the interfacial current data in Figure 1a well. This good agreement demonstrates the surface passivation dominated electrochemical oxidation of the carbon overcoat is an important mechanism at the high sliding speed interface. In a head-disk interface both the head and disk have a carbon overcoat. The head is coated with FCAC (field cathodic arc carbon) carbon on top which is more resistant than the nitrogenated carbon present on the disk \cite{WeiJAP88,KhunPCCS09,DwivediSR15}. The voltage polarity solely determines which surface is undergoing electrochemical oxidation. The inset of Figure 1a shows the associated net charge transfer in the electrochemical reaction for the negative and positive voltage. Charge associated mass transfer is driven more by the  negative voltage. We associate it with electrochemical oxidation on the disk as nitrogenated disk carbon is tribochemically less stable than FCAC head carbon. The associated wear or weight loss due to electrochemical oxidation is given by: 
\begin{equation}
Weight loss=\frac{qM}{4F}
\label{eq:2}
\end{equation}
where $q$ is the integrated total charge transfer across the interface, $M$ is the molecular weight, and $F$ is the Faraday constant. The factor of four assumes the number of electrons transferred in the head overcoat reaction is similar to the fuel cell oxidation case. Typically the area of contact on the head overcoat due to the heater bulge is around 10 $\mu m^{2}$. For positive voltage, Eq. 2 estimates the weight loss of head  overcoat to 8.75.10$^{-18}$ kg which corresponds to a wear depth of around 0.3 $nm$. This is in excellent agreement with the measured wear depth  (see red dots in Figure 1b), thus confirming the electrochemical oxidation of carbon overcoat as the dominant wear mechanism. For negative voltages, no head overcoat wear is observed (see blue dots in Figure 1b). Here the carbon overcoat on the disk undergoes electrochemical wear, and the head carbon remains intact, and leading to almost no wear under the same loading force conditions. The comparative volumetric loss for two polarities is more for positive bias, which results in high wear volume on disk overcoat. However, the overcoat wear on disk is diluted over a large area leading to a negligible wear depth in comparison to a localized wear on head where it could be quantified using AFM, SEM and Auger (see supplementary Figure S3 and Figure S4) \cite{JuangIEEE11,SuIEEE11,ChenTL14}.

Figure 2a shows the head overcoat wear as a function of sliding distance on the same head-disk interface. No head overcoat wear is observed for the first 1400 $m$ sliding distance under an applied bias of -1 $V.$ As the polarity is reversed to +1 $V,$ the wear rate increases sharply. This behavior exemplifies clearly that the voltage leads to asymmetric wear on head overcoat.         

\subsection*{Environmental effects on electrochemical wear} 
To gain further insight in the involved electrochemical process, we now turn to measurements in a controlled environment. The measurements are performed in an enclosed humidity controlled continuous flow set-up (as shown in Figure 3a). The chamber is connected to high purity gases (Nitrogen or dry Air) and the percentage of oxygen and humidity is monitored using gas sensors placed inside. 

Figure 2b shows the head overcoat wear depth as a function of head and disk interfacial voltage under atmospheric nitrogen condition with different relative humidity. The head overcoat undergoes the same wear cycle at each interfacial voltage and environment condition. The head overcoat wear depth is measured on an unused and pristine location on a disk after sliding on it for  3000 $m$ in intimate contact. It shows the importance of humidity on the head overcoat wear profile. At high relative humidity(RH) of 45$\%$, the head overcoat wear is not symmetric to the interfacial voltage. Positive voltage leads to high overcoat wear on the head, and negative voltage significantly suppresses the head overcoat wear. It is consistent with the overcoat wear as observed in Figure 1b. The impact of interfacial voltage polarity on the head overcoat reduces as the humidity content in the environment decreases.  At low RH of 7$\%$ the overcoat wear is very high and insensitive to voltage polarity. We attribute the high wear to the  slower tribochemical passivation of the dangling -OH bond on the surface at low humidity conditions \cite{Marino11Lagmuir}. The interfering surfaces are sliding at high speeds (almost six orders of magnitude faster than AFM), potentially leading to high flash temperatures and posing a greater challenge to stability of carbon under dry conditions.      

%%******************************************************************************************************
\begin{figure*}[htbp]
\begin{center}
\includegraphics[width=7 in]{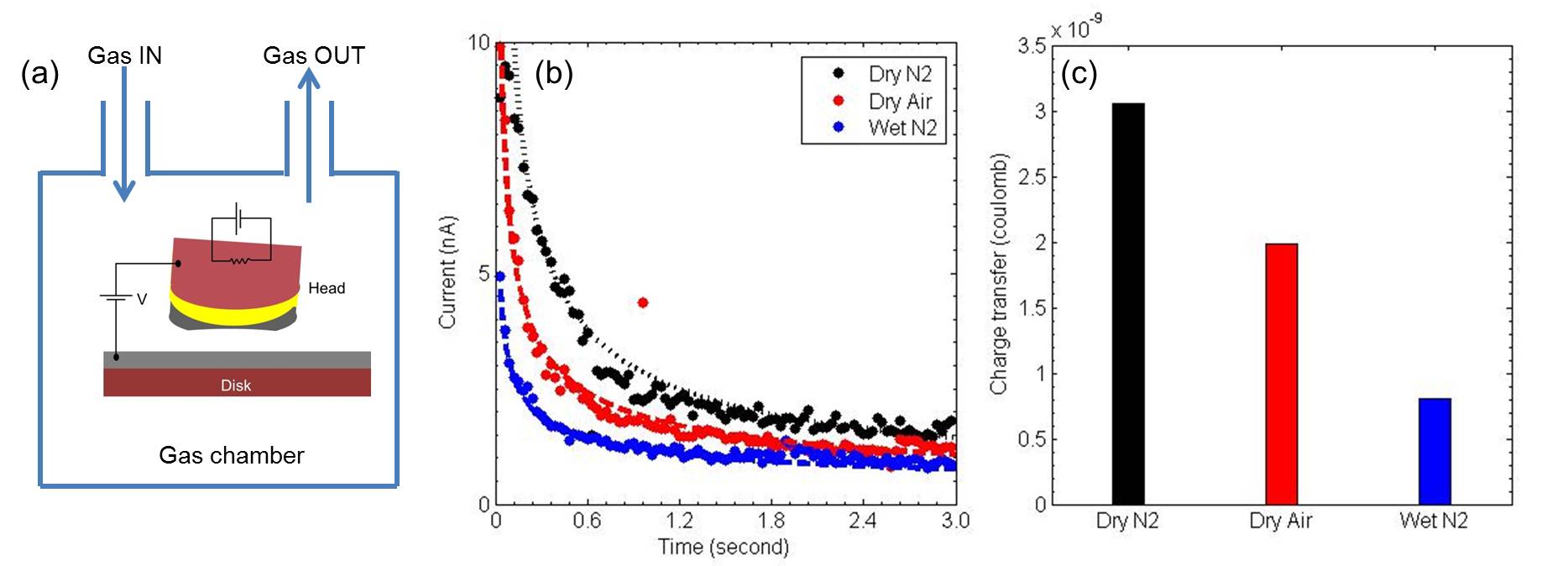}
\end{center}
\caption {\textbf{Environment control:} (a) Shows the set-up schematic of the high sliding speed head-disk interface in an environmental control chamber. (b) Black, Red and Blue dots represent the rapid decay of the interfacial current in dry Nitrogen, dry Air and wet Nitrogen environment conditions. The measurements are done at a normal load of 3.75 mN with -1.3 V applied on the head overcoat. Dashed line is fit to Eq. 1. (c) Shows the associated charge transfer for the three environment conditions.}
    \label{fig:3}
\end{figure*}
%******************************************************************************************************

To understand the complex electrochemical reaction, we investigate the effect of environmental conditions on the passivation process. The involved electrochemical reaction involves two competing parallel reactions: one to oxidize the carbon to produce carbon dioxide CO$_{2}$ leading to current in the circuit and another to passivate the surface leading to reduction in current. Figure 3b shows the interfacial current time decay at the head disk interface at -1.3 $V$ applied voltage on the head under a normal load of 3.75 $mN$ for three environmental conditions: dry nitrogen (black), dry air (red) and wet nitrogen (blue). The two dry conditions have RH<6$\%$ and wet air has RH=40$\%$. Compared to the two dry (nitrogen and air) conditions, the current under the wet nitrogen condition starts at lower value and saturates earlier, demonstrating the effectiveness of water in passivating the surface. Among the  two dry case (nitrogen and air), air has 20$\%$ atmospheric oxygen. As observed, molecular oxygen also contributes to surface passivation. Figure 3c shows the integrated charge transfer across the interface for the three environmental conditions. It shows that both humidity and molecular oxygen passivate the carbon overcoat with the former being more effective. 

\subsection*{Friction under electrochemical wear} 

In addition to an in-situ monitoring of wear and interfacial current, a measurement of the frictional properties is also desired to gain further insight into surface modification during sliding. The friction force between the head and the disk is also measured under a normal load condition. A calibrated strain gauge was instrumented to measure the friction force (in the sliding direction) between the head and the disk interface. The frictional force at an interface is written as, $F = \mu L + F_A$, where $F$ is the total friction at the interface, $\mu$ the non-dimensional coefficient of friction, $L$ the applied normal load, and $F_{A}$ the adhesion component of friction \cite{Bowden,MoNature,BermanTL98}. Figure 4 shows the delta friction force as a function of normal load when applied -1 $V$ and +1 $V$ on the head overcoat. Here, we have subtracted the friction contribution at low load (=1.25 $mN$) from all other load conditions to measure delta friction force. Dashed line is the linear fit with slope being the coefficient of friction. For positive voltage cycle on head, the friction coefficient is 0.4 in comparison to 0.2 for negative voltage cycle on the head. This further demonstrates the importance of the electrochemical activity on the carbon overcoat during sliding is determining the friction properties and long term durability of the overcoat.

%******************************************************************************************************
\begin{figure}[htbp]
\begin{center}
\includegraphics[width=3.4 in]{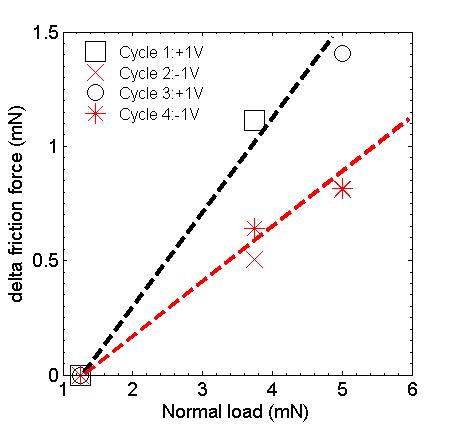}
\end{center}
\caption {\textbf{Frictional properties under an applied voltage:} Shows the delta friction force as a function of normal load under a repetitive cycle of applied +1 V and -1 V on head. Dashed line is the linear fit with slope being the friction coefficient.}
    \label{fig:4}
\end{figure}
%******************************************************************************************************

\subsection*{Chemical marking of contact location} 

%******************************************************************************************************
\begin{figure*}[htbp]
\begin{center}
\includegraphics[width=7 in]{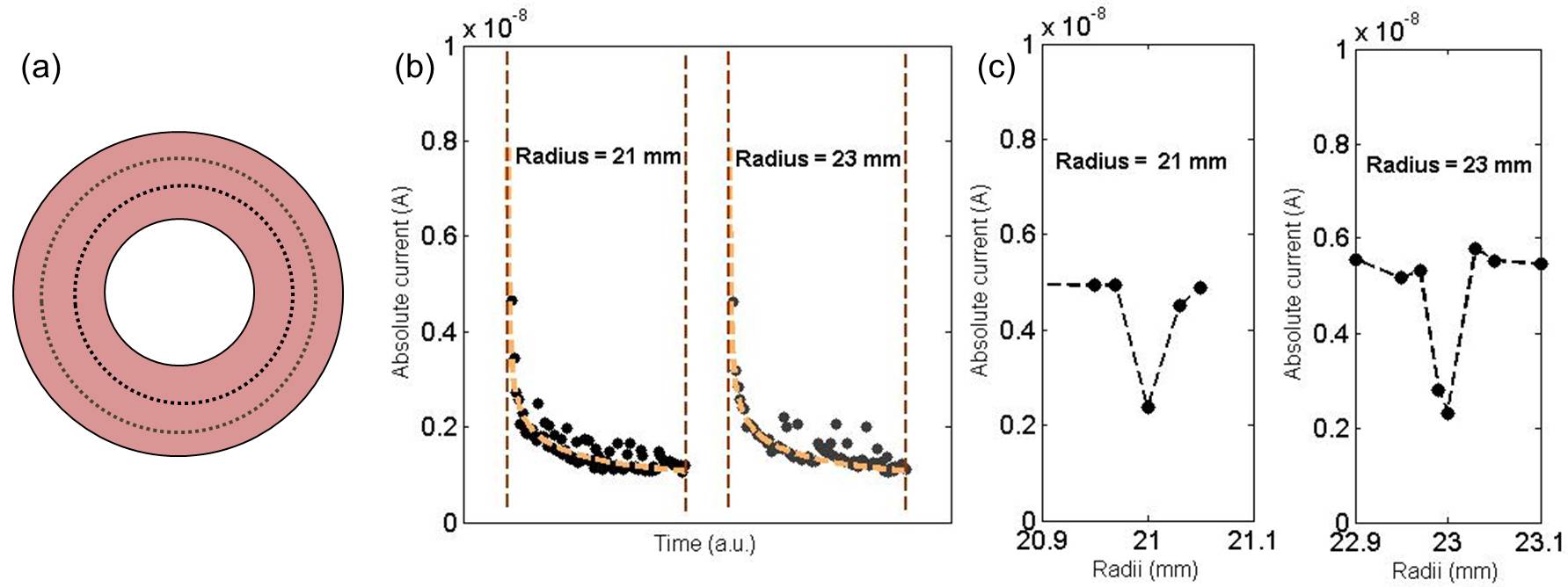}
\end{center}
\caption {\textbf{Chemical marking of contact location:} (a) Cartoon of disk depicting the intentional contact at two location, Radius =21 mm and 23 mm. (b) Marking: dots represent the passivation of carbon overcoat at two locations. (c) Detection: shows the interfacial current as probed by the same head at different tracks.}
    \label{fig:5}
\end{figure*}
%******************************************************************************************************
As a practical application of surface passivation, it was used to chemically mark the disk overcoat surface. Intentional contacts at two distinct tracks (radius 21 $mm$ and 23 $mm$ on disk) were made for 3 seconds under a  normal load of 3.75 $mN$ at the head disk interface. The head is held at -1 $V$ with respect to the disk. Figure 5b shows the decay in the interfacial current with the same head at two distinct tracks. Once the head passivates the first track on disk, the initial current on the new track is similar to the initial current of the previously passivated track. This further shows that for negative head voltages, the surface passivation dominated the electrochemical activity on the disk leaving the head in pristine condition. We used the same head to scan probe the electrical conductivity of the complete disk. Figure 5c shows the interfacial current as probed by the same head at different tracks. The electrical current on the passivated track is found to be significantly lower than the untreated area of the disk surface. We believe that this surface passivation of the carbon overcoat can have significant applications for high speed lithography. It is worth mentioning that recently AFM has been used to perform similar surface passivation of graphitic surfaces but AFM operates at typically six orders of magnitude slower sliding speed \cite{MordukhovichTL11}. Chemical analysis of the oxide formed on carbon overcoat is still missing and requires more work.

\section*{Discussion}

In summary, we have outlined a quantitative analysis of voltage assisted nanoscale electrochemical wear on carbon overcoat at high sliding speed interfaces. At high sliding speeds, in-situ measurements were performed of the interfacial current and the associated wear amount due to the electrochemical process. In addition, the effect of electrochemical activity on the interface is further quantified by measuring the friction force and the friction coefficient. It is found that the voltage assisted electrochemical activity greatly influences the interfacial wear and frictional properties. Positive voltage applied to the head leads to high wear on the head overcoat but no head overcoat wear was observed for negative applied voltage. As a useful application, we exploited the electrochemical passivation to mark the head-disk contact regions on the disk. The contact regions can be clearly identified by the associated conductivity variations of the surface. We believe that the observed voltage assisted asymmetric nanoscale wear will lead to additional experimental and simulation work, and will  help to understand precisely the chemical origin of the involved process.          

Our results are expected to have strong impact on fundamentally improving the carbon overcoat for various applications. The effect of interfacial voltage on the sliding interface is expected to be of great importance for understanding and improving the wear properties in nanoscale devices.     

\section*{Methods}

\subsection*{Sample preparation}
Disk: the rotating disk is a commercial 2.5" CoCrPt:oxide based hard disk media fabricated onto a glass substrate. Outermost thin film layer of hard disk media consist of 3 $nm$ amorphous nitrogenated carbon overcoat coated with a molecular layer of perfluoropolyether polymer lubricant ($\sim$ 1 $nm$ thick). 

Head: the head is a commercially available with read and write elements fabricated on a ceramic substrate. Similar to the disk, the head is also coated with a carbon overcoat with 1.4 $nm$ diamond like carbon on top of a 0.3 $nm$ Silicon based adhesive layer. The head surface is carefully etched (known as air bearing surface (ABS)) such that while flying on top of the disk an air lift force is generated that keeps it afloat in the nanometer distance over the disk.

\subsection*{Contact detection between the head and disk}
Contact between the  head  and  disk  is  monitored  using  a piezo-electric based  acoustic  emission  (AE)  sensor of the type PICO - 200-750 $kHz$. It  detects elastic propagating waves generated during the head-disk contact events \cite{BhushanIEEE03}.    
%******************************************************************************************************
\begin{figure}[htbp]
\begin{center}
\includegraphics[width=3.4 in]{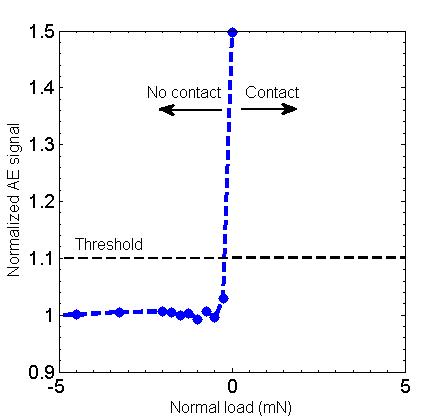}
\end{center}
\caption {\textbf{Contact detection:} Shows the normalized acoustic emission signal as a function of normal load between the head and the disk.}
    \label{fig:6}
\end{figure}
%******************************************************************************************************
Figure 6 shows a typical contact detection between the head and the disk. Vertical clearance between the head and disk is set using the embedded micro-heater inside the head. For protruding head making a contact with disk, AE signal increases sharply compare to  non-contact condition.  

\subsection*{Interfacial current and friction measurements}
Electrical measurement: In all interfacial measurement, the head-disk interface is first voltage biased then the contact is made using the micro-heater. The voltage bias is done using a HP 3314A source and the corresponding interfacial current is measured using Agilent 4155C. Interfacial current measurement are done under a positive normal load condition. 

Friction measurement: Contact friction force is  measured  using  a calibrated strain  gauge  mounted  at  the  end  of  the  suspension. The strain gauge signal is measured and amplified using a Vishay 2311 signal conditioning amplifier.

\section*{Acknowledgements}

The authors thank O. Ruiz for air bearing simulations, and S. Canchi, A., Murthy, N. Wang and V. Sharma for technical assistance. 

\section*{Author contributions statement}

S.R. and E.S. designed the experiments with contribution from B.M.. S.R. performed the experiments and analyzed the data. S.R., E.S. and B.M. wrote the manuscript.

\section*{Additional information}

The authors declare they have no competing financial interests.


\begin{thebibliography}{1}


\bibitem{Dowson}  D.H. Buckley,  {\textit{Surface Effects in Adhesion, Friction, Wear, and Lubrication} 429-508 (Elsevier, Amsterdam, 1981).}

\bibitem{Wimmer03} Wimmer, M. A., Sprecher, C., Hauert, R., Tager, G. and Fischer, A. 
Tribochemical reaction on metal-on-metal hip joint bearings - A comparison between in-vitro and in-vivo results.
{\textit{Wear} $\textbf{255}$, 1007-1014 (2003).}

\bibitem{Mosey05Science} Mosey, N. J., Muser, M. H. and Woo, TK
Molecular mechanisms for the functionality of lubricant additives.
{\textit{Science} $\textbf{307}$, 1612-1615 (2005).}

\bibitem{Bhushan94Nature} Bhushan, B., Israelachvili J. N. and Landman, U.
Nanotribology: friction, wear and lubrication at the atomic scale.
{\textit{Nature} $\textbf{374}$, 607-616 (1994).}

%\bibitem{Gosvami15Science} Gosvami, N. N., Bares, J. A. , Mangolini, F. , Konicek, A. R. , Yablon, D. G. and Carpick, R. W.  
\bibitem{Gosvami15Science} Gosvami, N. N. et al.  
Mechanisms of antiwear tribofilm growth revealed in situ by single-asperity sliding contacts.
{\textit{Science} $\textbf{348}$, 102-106 (2015).}

\bibitem{SukumarAPL15} Rajauria, S., Canchi, S. V., Schreck, E. and Marchon, B.
Nanoscale wear and kinetic friction between atomically smooth surfaces sliding at high speeds.
{\textit{Applied Physics Letters} $\textbf{106}$, 081604 (2015).}

\bibitem{ArchardJAP53} Archard, J. F.
Contact and rubbing of flat surfaces.
{\textit{Journal of Applied Physics} $\textbf{24}$, 981-988 (1953).}

\bibitem{JiaWear97} Jia, K. and Fischer, T.
Sliding wear of conventional and nanostructured cemented carbides.
{\textit{Wear} $\textbf{203}$, 310-318 (1997).}

\bibitem{ChungTL03} Chung, K. and Kim, D.
Fundamental investigation of micro wear rate using an atomic force microscope.
{\textit{Tribology Letter} $\textbf{15}$, 135-144 (2003).}



\bibitem{GneccoPRL02} Gnecco, E., Bennewitz, R.  and Meyer, E. 
Abrasive wear on the atomic scale.
{\textit{Phys. Rev. Lett.} $\textbf{88}$, 215501 (2002).}

\bibitem{GotsmannPRL08} Gotsmann, B. and Lantz, M. 
Atomistic wear in a single asperity sliding contact.
{\textit{Phys. Rev. Lett.} $\textbf{101}$, 125501 (2008).}

%\bibitem{BhaskaranNatureNano10} Bhaskaran, H., Gotsmann, B., Sebastian, A., Drechsler, U., Lantz, M. A., Despont, M., %Jaroenapibal, P., Carpick, R. W., Chen, Y. and Sridharan K.
\bibitem{BhaskaranNatureNano10} Bhaskaran, H. et al.
Ultralow nanoscale wear through atom-by-atom attrition in silicon-containing diamond-like carbon.
{\textit{Nature Nanotechnology} $\textbf{5}$, 181 (2010).}

\bibitem{JacobsNatureNano13} Jacobs, T. D. B. and Carpick, R. W.
Nanoscale wear as a stress-assisted chemical reaction.
{\textit{Nature Nanotechnology} $\textbf{8}$, 108-112 (2013).}

\bibitem{HanggiRMP90} Hanggi, P., Talkner, P. and Borkovec M.
Reaction-rate theory: fifty years after Kramers.
{\textit{Review of Modern Physics} $\textbf{62}$, 251 (1990).}




\bibitem{Robertson02} Robertson, J.
Diamond-like amorphous carbon.
{\textit{Materials Science and Engineering: R: Reports} $\textbf{37}$, 129-281 (2002).}

\bibitem{MarchonCarbon88} Marchon, B., Carrazza, J., Heinemann, H. and Somorjai, G. A.
TPD and XPS studies of O$_{2}$, CO$_{2}$ and H$_{2}$O adsorption on clean polycrystalline graphite.
{\textit{Carbon} $\textbf{26}$, 507-514 (1988).}

\bibitem{Marchon90IEEE} Marchon, B., Heiman, N. and Khan, M. R.
Evidence for tribochemical wear on amorphous carbon thin films.
{\textit{IEEE Transactions on Magnetics} $\textbf{26}$, 168-170 (1990).}

\bibitem{StromASME91} Strom, B. D., Bogy, D. B., Bhatia, C. S. and Bhushan B. 
Tribochemical effects of various gases and water vapor on thin film magnetic disks with carbon overcoats.
{\textit{Journal of Tribology-Transactions of the ASME} $\textbf{113}$, 689-693 (1991).}

\bibitem{DaiIEEE03} Dai, Q., Yen, B. K., White, R. L., Peterson, P. J. and Marchon, B.
Toward an understanding of overcoat corrosion protection.
{\textit{IEEE Transactions on Magnetics} $\textbf{39}$, 2450-2452 (2003).}

%\bibitem{KonicekPRB08} Konicek, A. R., Grierson, D. S., Sumant, A. V.,  Friedmann, T. A., Sullivan, J. P., Gilbert, P. U. P. A. , Sawyer, %W. G. and Carpick R. W.
\bibitem{KonicekPRB08} Konicek, A. R. et al.
Influence of surface passivation on the friction and wear behavior of ultrananocrystalline diamond and tetrahedral amorphous carbon thin films.
{\textit{Physical Review B} $\textbf{85}$, 155448 (2008).}

%\bibitem{Marino11Lagmuir} Marino, M. J., Hsiao, E., Chen, Y., Eryilmaz, O. L., Erdemir, A. and Kim, S. H.
\bibitem{Marino11Lagmuir} Marino, M. J. et al.
Understanding run-in behavior of diamond-like carbon friction and preventing diamond-like carbon wear in humid air.
{\textit{Langmuir} $\textbf{27}$, 12702–12708 (2011).}



\bibitem{KinoshitaCarbon73} Kinoshita, K. and Bett, J.
Electrochemical oxidation of carbon black in concentrated phosphoric acid at 135$^{o}$C.
{\textit{Carbon} $\textbf{11}$, 237-247 (1973).}

\bibitem{GallagheraPCCS09} Gallagher, K. G. and Fuller, T. F.
Kinetic model of the electrochemical oxidation of graphitic carbon in acidic environments.
{\textit{Physical Chemistry Chemical Physics } $\textbf{11}$, 11557-11567 (2009).}

\bibitem{ZilibottiPRB09} Zilibotti, G.,  Righi, M. C. and Ferrario M. 
Ab initio study on the surface chemistry and nanotribological properties of passivated diamond surfaces.
{\textit{Physical Review B} $\textbf{79}$, 075420 (2009).}




\bibitem{Mate87PRL} Mate, C. M., McClelland, G. M., Erlandsson, R. and Chiang, S.
Atomic-scale friction of a tungsten tip on a graphite surface.
{\textit{Phys. Rev. Lett.} $\textbf{59}$ 1942 (1987).}

\bibitem{Robertson01TFS} Robertson, J
Ultrathin carbon coatings for magnetic storage technology.
{\textit{Thin Film Solids} $\textbf{383}$, 81-88 (2001).}

\bibitem{Ferrari04Surface} Ferrari, A. C.
Diamond-like carbon for magnetic storage disks.
{\textit{Surface and Coatings Technology} $\textbf{180-181}$ 190-206 (2004).}





\bibitem{SuhTL06} Suh, A. Y., Mate, C. M., Payne, R. N. and Polycarpou, A. A.
Experimental and theoretical evaluation of friction at contacting magnetic storage slider-disk interfaces.
{\textit{Tribology Letters} $\textbf{23}$, 177-190 (2006).}




%\bibitem{cml}  CMLAir software available at: http://cml.berkeley.edu/ (Accessed: 12th January 2016).
\bibitem{cml}  CMLAir 8.3 (2015): CMLAir Bearing Design Program. URL http://cml.berkeley.edu/ (Accessed: 12th January 2016).



\bibitem{ZhengTL10}  Zheng, J. and Bogy, D.B
Investigation of flying-height stability of thermal fly-height control sliders in lubricant or solid contact with roughness.
{\textit{Tribology Letters} $\textbf{38}$, 283-289 (2010).}





\bibitem{ZengIEEE11}  Zeng, Q., Yang, C.-H.. Ka, S. and Cha, E.
An experimental and simulation study of touchdown dynamics.
{\textit{IEEE Transactions on Magnetics} $\textbf{47}$, 3433-3436 (2011).}


\bibitem{CanchiAT12} Canchi, S. V., Bogy, D. V., Wang, R. H. and Murthy A. N.
Parametric Investigations at the Head-Disk Interface of Thermal Fly-Height Control Sliders in Contact.
{\textit{Advances in Tribology} $\textbf{2012}$, 303071 (2012).}




\bibitem{ChenTL14} Chen, Y.-K., Murthy, A. N., Pit, R. and Bogy, D. B.
Angstrom scale wear of the air-bearing sliders in hard disk drives.
{\textit{Tribology Letters} $\textbf{54}$, 273-278 (2014).}





\bibitem{BhushanIEEE03}  Bhushan, B., Wu, Y. and Tambe, N. S.
Sliding contact energy measurement using a calibrated acoustic emission transducer.
{\textit{IEEE Transactions on Magnetics} $\textbf{39}$, 881-887 (2003).}





%\bibitem{Oscar} Ruiz, O.
%Air bearing simulation using HGST internal code.
%{\textit{Unpublished}.}


\bibitem{WeiJAP88} Wei, B., Zhang, B. and Johnson, K. E.
Nitrogen-induced modifications in microstructure and wear durability of ultrathin amorphous-carbon films.
{\textit{Journal of Applied Physics} $\textbf{83}$, 2491-2499 (1988).}


\bibitem{KhunPCCS09} Khun, N. W., Liu, E. and Krishna, M. D.
Structure, adhesive strength and electrochemical performance of nitrogen doped diamond-like Carbon thin films deposited via DC magnetron sputtering.
{\textit{Journal of Nanoscience and Nanotechnology} $\textbf{10}$, 4752-4757 (2010).}

%\bibitem{DwivediSR15} Dwivedi, N., Yeo, R. J., Satyanarayana, N., Kundu, S., Tripathy S. and Bhatia C. S.
\bibitem{DwivediSR15} Dwivedi, N. et al.
Understanding the role of nitrogen in plasma-assisted surface modification of magnetic recording media with and without ultrathin carbon overcoats.
{\textit{Scientific Reports} $\textbf{5}$, 7772 (2015).}




\bibitem{JuangIEEE11} Juang, J.-Y., Forrest, J. and Huang, F.-Y.
Magnetic head protrusion profiles and wear pattern of thermal flying-height control sliders with different heater designs.
{\textit{IEEE Transactions on Magnetics} $\textbf{47}$, 3437-3340 (2011).}

\bibitem{SuIEEE11} Su, L. et al.
Tribological and dynamic study of head disk interface at sub 1-nm clearance.
{\textit{IEEE Transactions on Magnetics} $\textbf{47}$, 111-116 (2011).}







\bibitem{Bowden}  Bowden, F.P. and Tabor D.  {\textit{The friction and Lubrication of Solids} 90-98 (Claredon, 1950).}

\bibitem{MoNature} Mo, Y. F., Turner, K. T. and Szlufarska, I.
Friction laws at the nanoscale.
{\textit{Nature} $\textbf{457}$, 1116-1119 (2009).}

\bibitem{BermanTL98} Berman, A., Drummond, C. and Israelachvili, J.
Amontons' law at the molecular level.
{\textit{Tribology Letters} $\textbf{4}$, 95-101 (1998).}



\bibitem{MordukhovichTL11}  Garcia, R., Knoll	A. W. and Riedo E.
Advanced scanning probe lithography.
{\textit{Nature Nanotechnology} $\textbf{9}$, 577–587 (2014).}


\end{thebibliography}
\end{document}